# The WKY-Haq Oscillator as Power Source of Inductive Power Transfer.

**Abdul-Wahab A. M. Adam***

*University of Benghazi, Faculty of Science, Department of Physics.*

**abdulwahab.adam@uob.edu.ly* ,   00218925533033

**Khalil Ismaiel Hashim[2]**

*University of Benghazi, Faculty of Science, Department of Physics*

khalilismaiel@gmail.com

**Dr. Yousef O. khazmi[3]**

*University of Benghazi, Faculty of Science, Department of Physics*

Yousefok@gmail.com

**Abstract.**

This paper is concerned with the wireless power transmission system based on inductance known as inductive power transfer (IPT), We have introduced a new oscillator called WKY-Haq, with an approximate mathematical relationship to adjust its frequency that was obtained experimentally in the electronics lab in physics department in the university of Benghazi. The WKY-Haq oscillator is a strong oscillator for operating the IPT system at low frequency 77.66 kHz with an excellent efficiency using the series-series (SS) Topology. Only the presence of a larger number of turns in the receiver can greatly improve the efficiency.

## I. Introduction

The wireless power transfer (WPT) is a transfer of the electric energy across an air gap without any physical contact. Which recently appeared as the wireless charging [1-4]. Despite all this interest in the recent years in wireless power transmission, this system is not new, as it dates back to 1893 when Tesla did his experiments and it had been proved that it was not impossible to achieve that transfer [2]. However, the development that happened to the devices which become of small size and also the wireless communications that made the devices as the mobiles easy to navigate, that's what made the WPT gain great momentum in recent years [1-4]. The WPT can be performed in far field under radiative system by electromagnetic radiation [5-8], also in near field by electric or magnetic field under reactive system, this study focuses on inductive power transfer (IPT) which depends on the electromagnetic inductance that classified under the reactive magnetic field (magnetic near field MNF), [9-12], also there are two methods under MNF, the magnetic resonant coupling (MRC) which depends on the resonance [13-15] and the magneto dynamic coupling (MDC) using permanent magnet [16,17]. On this basis, the energy transfer is classified according to the separation distance between the transmitter and the receiver, and then according to the adopted mechanism, radiative (electromagnetic) or reactive (magnetic or electric) field. The IPT system is considered the most effective and safest, which make it under study in several fields, and the most important field is the medical field in biomedical devices, for example the pacemaker which is used to set the heart rhythm, which contains a battery that has a limited lifetime and requires changing and sometimes the entire device needs to be replaced. An approach like IPT can solve these problems [10,12,18].

The IPT system is based on electromagnetic inductance that produces a magnetic field connecting two separated coils, with low frequency. The electromagnetic inductance discovered by Michael Faraday, who found that by changing a magnetic flux with time would produce a current in a closed loop of a wire [19,20].

According to Faraday's experiment, the production of a magnetic field through the coil requires a current that varies with time. A thought has been adopted to obtain an oscillating current source such as an inverter or an oscillator, which is defined as a DC to AC converter, its output is in the form of a wave such as sine, square, etc. with a frequency that can be controlled and tuned [21]. Then this AC current passes to a resonant circuit containing a capacitor and a coil that acts as an inductor that transmit magnetic flux lines across the air gap to another coil that acts as an inductor of the receiver that intercepts these lines and converts them into an electric current which passes through the inductor to another capacitor (resonant circuit of the receiver) to the load [9-12]. This processes is shown in Fig. (1):

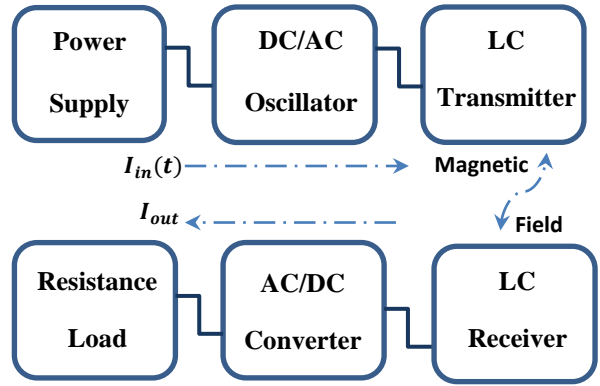

**Fig. (1): Diagram of the IPT System.**

In this paper we introduce a new oscillator of our design with a relationship of its frequency, this oscillator is effective for the use in the IPT system, using IC LM 7171 operational amplifier. We have given this oscillator the name WKY-Haq. The abbreviation of the names of those in charge of the project, Wahab, Dr. Khalil, and Dr. Youssef (WKY), and Haq attributed to Dr. Shams Al-Haq, one of the pillars of the physics department at the university of Benghazi, in his honor.

## II. Experiential Work

At the beginning of this work, the theory of the oscillator will be studied and a mathematical relationship will be found through which the frequency of the oscillator can be obtained by estimating the values of the parts of the oscillator. After the oscillator is equipped, it can be connected to the transmitter's resonant circuit and prepare the IPT system to work with our new oscillator.

### 2.1 The WKY-Haq Oscillator.

The oscillator is defined as an electronic circuit that converts DC current into AC current without an input signal [22]. In the initial setting of our oscillator, a Wien bridge was used with the LM7171 OP AMP and it was successfully tested. It was found through experimentation that the output and range of frequencies were appropriate, but its efficiency was not at the level required for wireless power transmission [23]. Then, we improved Wien bridge oscillator to obtain a new effective oscillator with less parts and good efficiency, that oscillator is given the name WKY-Haq oscillator as shown in Fig.(2):

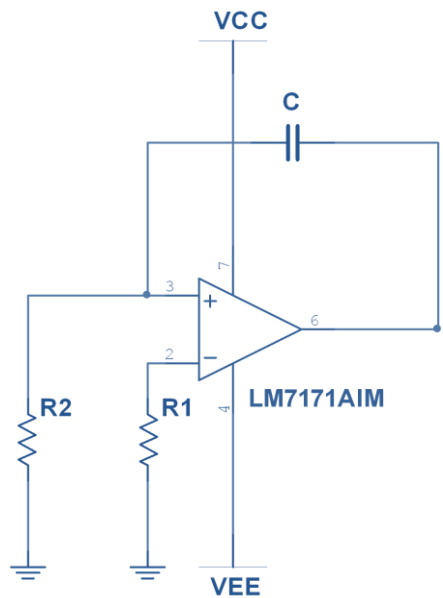

**Fig(2): The WKY-Haq Oscillator**

The circuit is similar in its structure to the integrator op amp, but it does not need an input signal and its output is completely different from the integrator op amp. it only worked with 7171 efficiently, although we tried it on some other op amps., it didn't work. The WKY-Haq oscillator was connected to the resonant circuit of the transmitter in order to test it with the receiver circuit for choosing the appropriate topology for this oscillator from the four topologies of transmitter and receiver resonant circuits. The resistors and capacitor values of the oscillator can be changed to get the best output that enhances the power delivered to the load, as stated in the IPT theory.

## 2.2 Tuning of WKY-Haq Oscillator.

The oscillator consists of two resistors and one capacitor, It was connected in the lab to a 12V and 1A power supply, with two equal value of resistors (can be used two capacitors) were used as a voltage divider to get +6V and -6V and a center-tap point between the two resistors. The waveform of this oscillator is shown in Fig.(3).

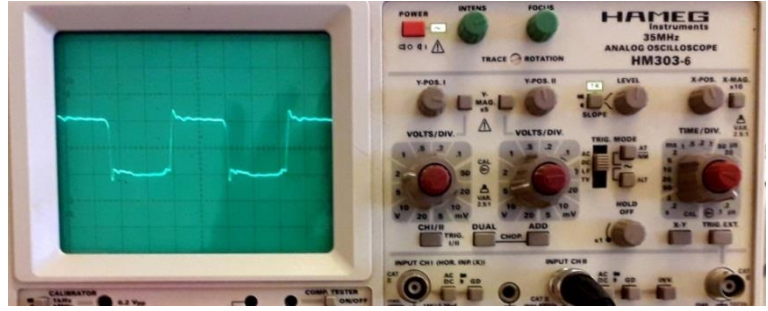

**Fig.(3): Waveform of WKY-Haq oscillator.**

Fig.(3), It can be noted that the oscillator waveform is square wave, which means that it has two time periods $t_1$ of the top part and $t_2$ of the bottom part, adding $t_1$ and $t_2$ gives the total time of the wave and the frequency can be found, When we changed the value of $R_1$, we found that the two time periods were not affected, but this resistor is important in the regularity of the wave shape. The other thing we noticed is that the condition $R_2 \geq R_1$ for the oscillator must be satisfied. The effect of the second resistor $R_2$ has been studied and the results recorded as in the table (1).

First, the change of the time periods as a function of the second resistor has been studied while fixing the value of the capacitor at 100pF and the first resistor at 200 ohm, and the obtained results are recorded in the table (1). We noticed that when the value of the second resistor was increased the time periods were also increased. Mathematically $t_1 \propto R_2$ and $t_2 \propto R_2$. Also, by changing the value of the capacitor, we noticed a direct proportion between the time periods of the wave, which were increased with the increase in the value of the capacitor. Mathematically $t_1 \propto C$ and $t_2 \propto C$, it can be written mathematically as follows:

$$A_1 = \frac{RC}{t_1} \qquad (1)$$

$$A_2 = \frac{RC}{t_2} \qquad (2)$$

where $A_1$ and $A_2$ are the proportionality constants of the two time periods of the wave, which can be calculated when changing the second resistor value and fixing the capacitor at 100pF which was found to be the best value for the capacitor then we got the best wireless power transmission. The period of whole cycle is $T$, where :

$$T = t_1 + t_2 \qquad (3)$$

The frequency of the obtained wave can be calculated by:

$$f = \frac{1}{T} \tag{4}$$

**Table(1): effect of time period of WKY-Haq oscillator waveform.**

| $R_2$ Ω | $t_1$ μs | $A_1$ | $t_2$ μs | $A_2$ | $T$ μs | $f$ |
|---|---|---|---|---|---|---|
| 100 | <td colspan="6">Unstable because $R_2 < R_1$.</td> |||||||
| 220 | 0.18 | 0.12 | 0.22 | 0.10 | 0.40 | 2.5 MHz |
| 470 | 0.26 | 0.18 | 0.32 | 0.20 | 0.58 | 1.7 MHz |
| 1000 | 0.49 | 0.20 | 0.48 | 0.21 | 0.97 | 1.0 MHz |
| 2200 | 0.85 | 0.26 | 0.75 | 0.29 | 1.60 | 625 kHz |
| 4700 | 2.40 | 0.20 | 1.40 | 0.34 | 3.80 | 263 kHz |
| 10000 | 4.00 | 0.25 | 2.80 | 0.36 | 6.80 | 147 kHz |
| 15000 | 6.00 | 0.25 | 4.50 | 0.33 | 10.5 | 95.2 kHz |
| 20000 | <td colspan="6">Unstable</td> |||||||

from table(1) we can calculate the mean value of the first time constant $A_1$ which was found to be 0.21 and the second time constant $A_2$ was found to be 0.26. Since the two values are very close, a mean value for them can also be calculated and was given a symbol $A$, where for them $A = 0.235$. It is possible to find the values of the time periods by Eqs. (1) and (2):

$$t_1 = \frac{RC}{0.235} \tag{5}$$

And:

$$t_2 = \frac{RC}{0.235} \tag{6}$$

The two time periods are approximately equal, by applying the Eq. (3), from which the total time is:

$$T = \frac{2RC}{0.235} = 8.51 RC \tag{7}$$

and the generated frequency is calculated using Eq.4 to be:

$$f_c = \frac{1}{8.51RC} \tag{8}$$

This approximate frequency relationship enables us to estimate the required frequency in a range close to the actual frequency of the oscillator. In order to confirm Eq.(8), the value of the actual oscillator frequency measured in the laboratory $f_{osc}$ by multimeter (UNI-T 890D+) can be compared with the frequency mathematically calculated $f_C$ .

Table(2): Comparing between $f_{osc}$ and $f_C$

| C pF | $f_{osc}$ kHz | $f_C$ kHz |
|---|---|---|
| 60 | 203.6 | 196.6 |
| 100 | 126.6 | 118.0 |
| 220 | 56.74 | 53.61 |
| 294 | 46.00 | 40.12 |
| 440 | 31.25 | 26.81 |
| 500 | 23.22 | 23.59 |
| 700 | 17.83 | 16.85 |
| 950 | 14.57 | 12.42 |
| 1000 | 13.11 | 11.80 |
| 4000 | 03.60 | 02.95 |
| 10000 | 01.56 | 01.18 |

We notice from table (2) that the two values of the two frequencies almost equal. Fig.(4) shows the curves of the two frequencies are closed, that means the relationship (8) is very suitable for this oscillator.

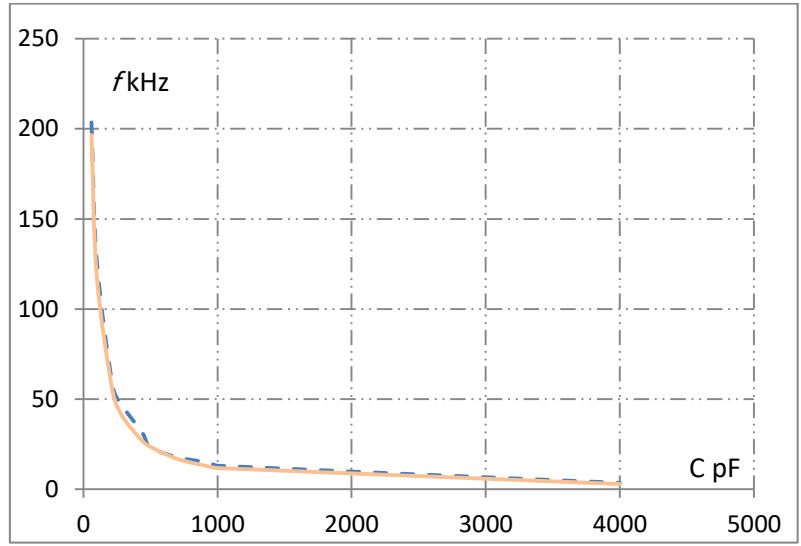

**Fig.(4): Comparing between $f_{osc}$ and $f_C$**

in Fig.(4) the blue line is $f_{osc}$ and the yellow line is $f_C$.

## 2.3 Operating IPT System with WKY-Haq Oscillator.

The WKY-Haq oscillator was connected to the resonant circuits of the transmitter and receiver and the system was tuned to ensure the highest energy transfer, as shown in Fig.(5). In this device we used aluminum coils equivalent to copper in diameter (0.5mm) and the number of turns of the receiver coil was greater than that of the transmitter; we noticed an improvement in energy transfer. The SS topology was found to be the best. We connected the oscillator to a push-pull circuit to enhance the current arriving in the transmitter's resonant circuit, and the transmission also improved greatly, and lead to a greater magnetic flux by the transmitter inductor, but heat was produced in the two transistors, one of which may be disrupted. It can be overcome by using a suitable type of capacitors in the resonant circuits to pass the current at an appropriate amount, which would eliminate heat and maximize the transfer power of the load, thus improving efficiency better than ceramic capacitors, also choosing the appropriate push and pull transistors for the amount of current supplied from the power source. Therefore, we used a power transistor in push pull circuit. The device was then ready to study as shown in Fig.(5).

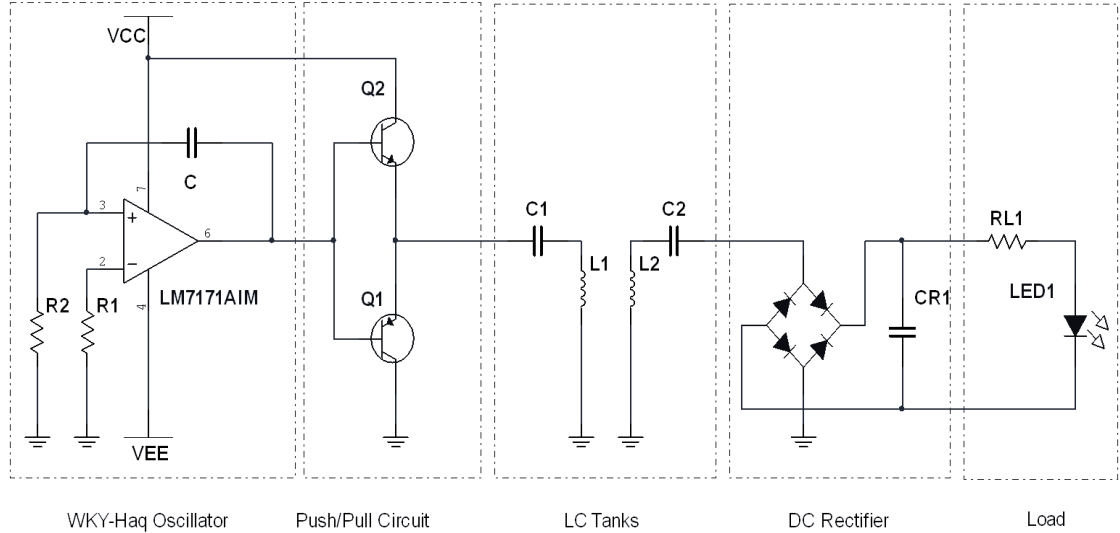

**Fig.(5): The IPT system with WKY-Haq oscillator as power source.**

The power transistors in the push-pull circuit have the ability to withstand the current passing through them and to reduce the heat that a normal transistor cannot handle. The values of the parts were as follows:

**Table(3): The values of the component of the IPT system.**

| Component | Value |
|---|---|
| $R_1$ | 500Ω |
| $R_2$ | 10KΩ |
| $C_{osc}$ | 100pF |
| $C_1$ | 150nF |
| $C_2$ | 46.63nF |
| $L_1$ | 26.31μH |
| $L_2$ | 81.83μH |

A 1cm separation distance between two inductors was fixed, and by adding a push-pull circuit to our oscillator that boosts the current, we were able to measure the output power of this circuit, the results were shown in table (4):

**Table (4): the resulting of IPT system with WKY-Haq oscillator**

| $f_1$kHz | $I_1$mA | $V_1$volt | $P_1$Watt | $f_2$kHz | $I_2$mA | $V_2$volt | $P_2$Watt | $\eta$ |
|---|---|---|---|---|---|---|---|---|
| 77.66 | 1000 | 6 | 6 | 77.89 | 207.4 | 19.05 | 3.95 | 65.8% |

where $P_1$ is the power of the oscillator output and $P_2$ is the power at the load. Improving efficiency of WPT can be studied by many ways, the frequency tracking to choose certain frequency which we consider the key of the best wireless power transmission. In this experiment, when supplied with a source of 12 volts and 1 amp, we obtained a power out of the oscillator in watts, where the power at the load was 3.95 watts in return for an efficiency equal to 65.8%, but the oscillator can be supplied with a higher power supply, when changing the value of the power source from 0 to the highest value of the voltage for 7171, we found that the IPT system with the same components mentioned in Table 7 starts working at a power source of 7 volts up to 30 volts, but the best performance was at 20 volts and 1.5 amps at a frequency of 75 kHz.

Note:

1. The measurements were made using the UNI-T multimeter type of UT890D+ for measuring current, voltage, and frequency, and the L/C meter type of LC200A for measuring capacitance and inductance.
2. $V_2$ in table (7) measured at both ends of the DC rectifier without load resistor.

## Conclusion

We have introduced a new oscillator that works with good efficiency in the IPT system approved in medical devices. The WKY-Haq oscillator works on the SS topology in an excellent way. Its performance can also be improved by connecting a greater number of turns in the receiver than in the transmitter. The best operating efficiency was at a frequency of 77.66 kHz, and this is excellent for safety standards, especially for use in the medical field for devices implanted inside the human body. Wireless power transmission can be improved through the use of a push-pull circuit. We recommend the use of better designs for inductors that have a coupling coefficient greater than 0.5, where the mutual inductance is greater than that in the simple designs for better efficiency.